\documentclass{article}
\usepackage{arxiv}
\usepackage[utf8]{inputenc}
\usepackage[colorinlistoftodos]{todonotes}
\usepackage{makecell}
\usepackage{subfigure}
\usepackage{url}
\usepackage[backend=bibtex, natbib=true, style=numeric-comp, sorting=none, isbn=false, url=false]{biblatex}
\newcommand{\ket}[1]{\ensuremath{|#1\rangle}}
\newcommand{\bra}[1]{\ensuremath{\langle#1|}}

\newcommand{\cmmnt}[1]{}

\definecolor{poop}{RGB}{100, 70, 20}
\definecolor{puke}{rgb}{0.6, 0.71, 0.0}

\title{Reinforcement learning for optimization of variational quantum circuit architectures}

\author{Mateusz Ostaszewski\\
	Institute of Theoretical and Applied Informatics, \\Polish Academy of Sciences, \\ Gliwice, Poland
	\\ \texttt{mm.ostaszewski@gmail.com}, 
	\And
	Lea M. Trenkwalder\\
	Institute for Theoretical Physics,\\ University of Innsbruck\\  Innsbruck, Austria\\
	\texttt{lea.trenkwalder@uibk.ac.at}, 
	\AND
	Wojciech Masarczyk \\
	Warsaw University of Technology, \\ Warsaw, Poland\\
	\texttt{wojciech.masarczyk@gmail.com}, 
	\AND
	Eleanor Scerri\\
	Leiden University,\\ Leiden, The Netherlands\\
	\texttt{scerri@lorentz.leidenuniv.nl}, 
	\And
	Vedran Dunjko\\
	Leiden University,\\ Leiden, The Netherlands\\
	\texttt{v.dunjko@liacs.leidenuniv.nl}}

\begin{document}
	\maketitle

	\begin{abstract}
		The study of Variational Quantum Eigensolvers (VQEs) has been in the spotlight in recent times as they may lead to real-world applications of  near term quantum devices.  
		However, their performance depends on the structure of the used variational ansatz, which requires balancing the depth and expressivity of the corresponding circuit.
		In recent years, various methods for VQE structure optimization have been introduced but capacities of machine learning to aid with this problem has not yet been fully investigated.
		In this work, we propose a reinforcement learning algorithm that autonomously explores the space of possible ans{\"a}tze, identifying economic circuits which still yield accurate ground energy estimates. The algorithm is intrinsically motivated, and it incrementally improves the accuracy of the result while minimizing the circuit depth.
		We showcase the performance or our algorithm on the problem of estimating the ground-state energy of lithium hydride (LiH).
		In this well-known benchmark problem, we achieve chemical accuracy, as well as state-of-the-art results in terms of circuit depth.
	\end{abstract}


	\section{Introduction}

	As we are entering the so called Noisy Intermediate Scale Quantum (NISQ)~\cite{preskill2018quantum} technology era, the search for more suitable algorithms under NISQ restrictions is becoming ever more important. A truly compatible NISQ application must first be amenable to architecture constraints and size limits. Furthermore, to minimize the adverse effects of gate errors and decoherence, it is important that the circuits are as gate-frugal, and as shallow as possible.

	Perhaps the most promising classes of such algorithms are based on variational circuit methods, with which we have high expectations when applied to problems in quantum chemistry.  
	A key problem in this field is the computing of ground state energies and low energy properties of chemical systems (the chemical structure problem).
	This problem is believed to be intractable in general, yet the quantum 
	Variational Quantum Eigensolver (VQE)~\cite{peruzzo2014variational} algorithm can provide solutions in regimes which lie beyond the reach of classical algorithms, while maintaining NISQ-friendly properties~\cite{kandala2017hardware,grimsley2019adapt}.

	VQE is a hybrid quantum-classical algorithm, where a parametrized quantum state is prepared on a quantum computer, the parameters of which are selected using classical optimization methods. 

	The objective is to prepare the state $| \psi(\vec{\theta})\rangle$ which can be used to approximate the ground state of a given Hamiltonian $H$ by the variational principle
	\begin{equation}
		E_{\min} \leq \min_{\vec{\theta}}(\bra{\psi(\vec{\theta})}H\ket{\psi(\vec{\theta})})~,
		\label{eq:vqe}
	\end{equation}
	where $E_{\min}$ is the true ground state energy of $H$. 
	The parametrized state is prepared by applying $U(\vec{\theta})$, 
	which is a parametrized quantum circuit (typically with a fixed architecture), where the angles $\vec{\theta} = (\theta_1 ... \theta_n)$ specify the rotation angles of the local unitary rotations present in the circuit.
	This circuit, known as the \textit{ansatz} is applied to an initial state $\ket{\psi_0}$, usually chosen to be the fiducial ``all zero'' state $\ket{0}$, to prepare the state $\ket{\psi(\vec{\theta})}=U(\vec{\theta})\ket{\psi_0}$.

	It is well established that the structure of the ansatz can dramatically influence the VQE's performance \cite{grimsley2019adapt, tang2019adapt}, 
	as the closeness of the estimated ground state to the true one depends on the state manifold accessible by the ansatz.
	Thus finding new architecture construction methods could lead to breakthroughs in quantum variational methods for chemistry (e.g. for strongly-correlated systems, for which standard ans\"{a}tze might fail), but also in other domains which utilize variational circuits such as machine learning and optimization \cite{dallaire2018gan, benedetti2019pqc, zhou2020qaoa, romero2021gan}. 
	
	Currently, the foremost ans\"{a}tze fall primarily in two classes: chemistry-inspired (e.g. the unitary coupled-cluster ansatz \cite{peruzzo2014variational, yung2014variational}) and hardware-inspired (e.g. the hardware efficient ansatz \cite{kandala2017he}). Architectures from both of these classes entail using a fixed architecture~\cite{peruzzo2014variational, yung2014variational,kandala2017he,benedetti2020hardwareefficient} determining the unitary $U(\vec{\theta})$, and hence the corresponding ansatz. The ansatz circuit is then usually decomposed into two-qubit CNOT and one-qubit rotation gates parametrized by $(\theta_1,\theta_2,\dots,\theta_n) \in [0,2\pi]^n$ to be optimized by a classical subroutine. However, the architecture itself can also be optimized. This results in a hard structure optimization problem, as it is a combinatorial optimization problem which must balance two competing factors. On the one hand the ansatz needs to be expressive enough to guarantee a good approximation of the ground state energy. On the other hand, the depth and size of the circuit needs to be controlled in order for the latter to be compatible with NISQ restrictions.
	
	Recently, the use of machine learning in problems of quantum computing has gained popularity. This is a very promising approach because of at least two reasons. Firstly, the use of problem-agnostic models reduces the need for expert knowledge and human input~\cite{sorensen2016exploring}, whilst providing good quality solutions~\cite{khairy2020learning,Ostaszewski2019approximation,ostaszewski2020geometrical,banchi2018modelling,dalgaard2020global}. Secondly, based on the success of deep learning towards solving difficult problems, such as protein folding \cite{senior2020improved}, 
	or significantly outperforming humans in a variety of tasks ~\cite{sorokin2020interferobot,silver2016mastering,ecoffet2021first}, it is a natural candidate for solving difficult problems such as the construction of quantum circuits.

	In this work, we propose a general optimization procedure for VQE based on deep reinforcement learning (RL) which is designed to yield quantum circuits that are both gate and depth efficient. 
	We supplement this RL approach with curriculum learning \cite{Elman1993, Bengio2009}, a powerful machine learning method for solving complex problems by leveraging the solutions of previously-solved simpler instances. Specifically, we introduce a feedback-driven curriculum learning method that autonomously adapts the complexity of the learning problem to the current performance of the learning algorithm.
	We apply our architecture to the well-established benchmarking problem of finding the ground state energy of the LiH molecule~\cite{csizmadia1964group,kandala2017hardware} and observe chemical accuracy while maintaining a low-depth quantum circuit and achieving state-of-the-art results in terms of gate efficiency.  Moreover, the proposed method is much broader than just for the ground state finding, it can be applied to any variational-circuit-based algorithm.
	
	The rest of the paper is organized as follows. In Section 2 we discuss the related works. In Section 3 we introduce our method with simplistic assumptions about the exact value of the ground state energy. At the end of this section we provide a method circumventing this unrealistic requirement. To ease the experiments, in Section 4 we present the results assuming the access to the exact value of energy. In Section 5 we provide the results for the full method that utilizes a rough proxy of energy obtained with classical methods. We conclude the paper with the discussion of possible future directions.

	\section{Related Work}

	The automatic construction of quantum circuits with the use of heuristics/machine learning methods is a topic that has recently been gaining more attention. In some lines of research, the focus are certain important sub-problems or special cases of the overall task, such as qubit routing \cite{herbert2018using}, gate synthesis and state preparation \cite{alam2019quantum}.
	In~\cite{pirhooshyaran2020quantum}, proposed approach is somewhat similar to ours in that the authors leave the process of constructing the architecture to the agent, while the optimization of its parameters is a factor independent of the agent, and is used to provide feedback for the agent. However, the authors use RL to construct parametrized quantum circuits for data classification.  
	In a more recent work \cite{fösel2021quantum}, the authors use RL to optimize the structure of square circuits in terms of their depth and number of quantum gates. As part of the action, the agent transforms the initial quantum circuit. As an example of a real application, the authors optimize the QAOA circuit to solve the MaxCut problem.
	
	In a related vein, the authors in \cite{Rattew2019ADN} use evolutionary algorithms to optimize the structure of VQE circuits for quantum chemistry and combinatorial optimization. The gate set in this algorithm is composed of one-qubit universal gates and two-qubit controlled universal gates, which after optimization procedure, are decomposed into CNOT gates and one-qubit universal rotation gates. The authors use three different mutation strategies: random appending of gates to each qubit in a circuit layer (with zero-initialised parameters), sequential optimization of each layer based on a random order, and removal of a random number of neighbouring gates in a sequence.
	The objective function used for the fitness evaluation includes the energy evaluated using the circuit corresponding to the genome, as well as two penalty terms, one for the number of gates used from the alphabet and and another specifically for the number of two-qubit gates. 
	Another evolutionary strategy was explored in~\cite{chivilikhin2020mog}, where a multi-objective genetic algorithm is used to optimize the structure of the VQE ansatz in the context of quantum chemistry problems. The topology of the circuits is optimized by a non-dominated sorting genetic algorithm while the parameters are globally tuned by Covariance Matrix Adaptation Evolution Strategy (CMA-ES) \cite{hansen2001}. The goal is to achieve high performance quantum circuits with as few two-qubit quantum gates as possible. 
	
	\section{Methods}
	
	In this section, we discuss our approach for constructing VQE ans{\"a}tze using RL. After briefly introducing the RL framework, we discuss the state and action representations, as well as the reward function used in this work. We then introduce a novel adaptation of curriculum learning applied to VQE circuit construction with different parameter optimization strategies.

	\subsection{Ansatz optimization as a reinforcement learning problem}
	
	To design shallow quantum circuits that estimate the energy within chemical accuracy, we utilize deep RL methods that combine methods from RL with deep neural networks used as function approximators. In RL the learning algorithm is referred to as an agent, which learns the correct, or at least nearly-optimal, behaviour by interacting with the environment and receiving a reward signal based on its performance. To specify a RL setting corresponding to a problem we wish to tackle, one needs to specify a set of all possible states $\mathcal{S}$ the agent can visit, the set of all possible actions $\mathcal{A}$ that agent can take and the set of rewards $\mathcal{R}$ that agent can obtain. The whole dynamics of the environment is governed by two operators responsible for assigning a probability of transition to a particular state given the current state and actions $T: (\mathcal{S} \times  \mathcal{A}) \times \mathcal{S} \rightarrow [0, 1]$ and a reward function $R: (\mathcal{S} \times  \mathcal{A}) \times \mathcal{S} \rightarrow \mathcal{R}$.
	
\begin{figure}
		\centering
		\includegraphics[width=\textwidth]{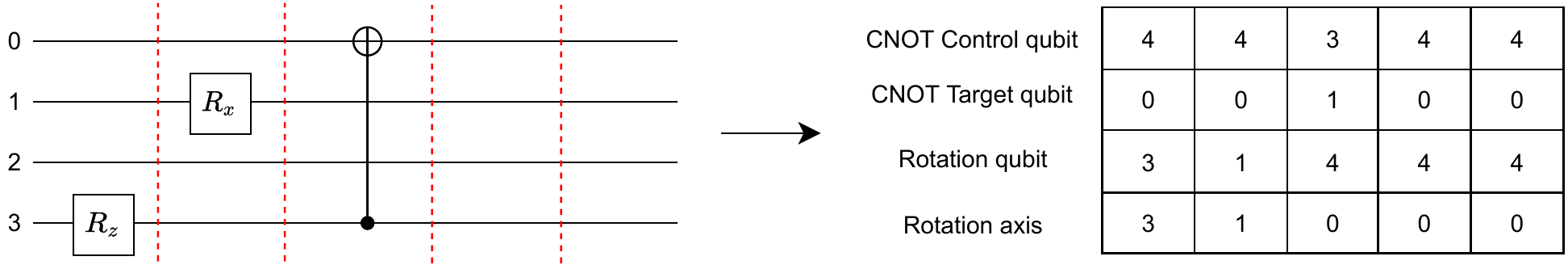}
		\caption{Example of state representation. In this example maximum length of the circuit ($L$) is set to 5, and the number of qubits to 4. Since we count qubits from 0, the lack of particular gate at each layer is represented with the maximum number of qubits, which in this case is 4, therefore the last two columns represent layers with identity operators. 
		}
		\label{fig:state_representation}
	\end{figure}

	To utilize the deep RL algorithms we design the representations of states and actions in a neural network-friendly form. Namely, each state is represented as an ordered list of layers that are composed of single quantum gates. This list uniquely defines the whole quantum circuit. 
	Therefore, the order of elements in the list describes the order of quantum gates in the circuit. For constructing the circuits, we use CNOT and one qubit rotation gates -- which are realizable on currently available quantum devices. CNOT gates are encoded by two values indicating position of \textit{control} and \textit{target} qubits (position is enumerated from 0). This way, the agent has full flexibility and in principle can insert CNOT gate between two arbitrary qubits (except for the same qubits). Rotation gates are encoded with respective positional number (starting from 0) indicating the qubit on which to apply the rotation, and the axis of rotation (\textit{i.e.}, the Pauli operator X, Y , Z generating the rotation which are represented as integers 1,2,3). To fully describe the rotation gate we need to specify the angle of the rotation, however we deliberately omit this parameter from the state representation. At the end of each state's representation we append the energy estimated for the state.
	In our approach, the user needs to define the maximum number of circuit layers ($L$).
	If a given circuit has fewer layers than $L$, the remaining ones are filled with identity operators (c.f. Fig.~\ref{fig:state_representation}).
	
	In our approach the agent constructs the circuit iteratively, starting from an empty list, and adding a single quantum gate at the end of the current circuit. This way the agent starts each episode with exactly the same conditions, \textit{i.e.} an empty state, avoiding an additional source of randomness tied to the initial state.

	Thus the set of all possible actions is equivalent to the set of all possible quantum gates that the agent can select (even the ones that do not make sense from the perspective of quantum circuit, \textit{i.e.} repetitions of rotation or CNOT gates). 
	The size of the set of all possible actions for the agent is then $3|Q| + 2{{|Q|}\choose{2}} = |Q|(|Q|+2)$, where $|Q|$ is the number of qubits, $3 |Q|$ is the number of single qubit gates ($|Q|$ qubits and 3 axes of rotation) and $2{{|Q|}\choose{2}}$ is the number of two qubit gates the agent can choose from. We use the simplest possible one-hot encoding for the actions.

	After appending each gate we evaluate current circuit's energy based on the following formula:
	\begin{equation}
		E_{t} = \bra{\psi(\vec{\theta})}H\ket{\psi(\vec{\theta})},
		\label{eq:energy}
	\end{equation}
	where $\vec{\theta}$ describes angles for the rotational gates and $t$ denotes timestep of current episode.
	Within a single episode, at each step the agent receives the reward according to the following formula:
	
	\begin{equation}
		R=\left\{
		\begin{array}{ll}
			5 & \textup{ if } E_{t} < \xi \\
			-5 & \textup{ if } t > L \\
			\max{(\frac{E_{t-1} - E_{t}}{E_{t-1}-E_{\min}},-1)} & \textup{otherwise} \\ 
		\end{array} ,
		\right.
		\label{eq:reward}
	\end{equation}

	where $\xi$ is a predefined threshold 
	of how close one wants to be to the exact energy. Further details on the choice of $\xi$ are discussed in Sec.~\ref{seq:curriculum}. Therefore, if the energy is below the threshold, a maximal reward of $+5$ is given, whereas the minimal reward of $-5$ is given if the circuit has more than $L$ layers. If the circuit neither satisfies the threshold $\xi$ nor does it go over the maximal number of layers $L$, an intermediate reward is given. This intermediate reward, $\frac{E_{t-1} - E_{t}}{E_{t-1}-E_{\min}}$ (which ideally would be equal to 1), is capped by $-1$ whenever the estimate at timestep $t$ is significantly worse than that at $t-1$ (recall that $E_{t-1}-E_{\min} \geq 0$)
	
	The extreme reward values $\pm 5$ are crucial for the performance of the agent. We hypothesize that larger extreme rewards facilitate the agent's learning of the correct architectures, as the discovery of desired circuits is highly rewarded (independent from the current energy difference). We include the term $\frac{E_{t-1} - E_{t}}{E_{t-1}-E_{\min}}$ in the reward function so as to avoid a highly sparse reward. Finally, the episode terminates if either the length of the circuit exceeds $L$ or the agent receives the maximum reward of 5.

	\subsection{Specification of the agent and the environment} 
	
	In this paper, we employ a Double Deep-Q network (DDQN) with an $\epsilon$-greedy policy and an ADAM optimizer. DDQN is an RL method that, similar to the standard Deep-Q network~\cite{mnih2013atari}, derives a policy from the Q-function approximated by a neural network. However, the overestimation bias is reduced in DDQN  by using two separate networks, one for Q-value evaluation and another for selecting actions, where the former is an earlier copy of the latter.
	
	According to Eq. \ref{eq:energy}, in order to estimate the energy obtained by a particular circuit one needs to determine the continuous values rotational gate angles. We use the well developed/standard methods for continuous optimization such as Constrained  Optimization  By  Linear  Approximation (COBYLA)~\cite{powell1994direct} and Rotosolve~\cite{ostaszewski2021structure}. Whilst we could use an agent for this purpose as well, we abstain from doing so for two main reasons.
	Firstly, in this work we aim at assessing the efficacy of RL for circuit structure optimization, rather than for parameter searching. The latter is investigated extensively in other literature and thus treated as a separate subroutine that can be solved by other, more developed, tools. The second reason is practical, as training the agent to reliably predict values of angles would result in significantly longer training periods.
	
	For the above reasons, our approach is hybrid: the agent learns how to, and is rewarded for, constructing particular circuits given set of parameters determined by an independent optimization algorithm. We apply the angle optimization subroutine only after the steps in which the agent appends a rotation gate to the circuit, thus eliminating the necessity of including the parameters in the state representation. The number of angles optimized at a given step is a hyperparameter in our method, the choice of which will be analyzed in Sec.~\ref{sec:Results}. In this work, we consider optimizing all angles at once (\emph{global strategy}), as well as optimizing a few angles at a time (\emph{local strategy}) . 
	These different experimental settings allow us to check whether the rough energy approximation obtained from only optimizing a subset of the circuit angles is sufficient to create sufficiently good circuits. 

	\subsection{Intrinsically motivated curriculum learning}\label{seq:curriculum}

	The goal of the algorithm is to find circuits that estimate the true ground state energy given within some threshold $\xi$ for a particular Hamiltonian. In this work, we choose this threshold to represent the chemical accuracy, i.e. get an energy that is close to the ground state energy by approximately 0.001 Hartee. 
	In the simplest case, in order to assess whether the proposed circuit passes the threshold one needs to compute the exact value of minimal energy. However, in practice this is often infeasible due to the computational requirements that grow exponentially with respect to number of qubits in the analyzed system. To circumvent this issue we propose a novel method that automatically creates a sequence of tasks for the agent and simultaneously avoids the need of computing minimal energy.

	Even in cases where the exact value of minimum energy is known, setting the value of $\xi$ directly to chemical precision from the beginning of the training may result in the agent failing due to poor exploration. Namely, as the threshold is significantly stringent, the agent might not be able to reach circuits which satisfy this threshold and hence obtain the maximal reward value (c.f. Eq.~\ref{eq:reward}). One of the well-studied solutions to that problem, known as curriculum learning~\cite{Elman1993, Bengio2009}, requires the specification of a sequence of tasks with ever-increasing level of difficulty. This way the agent starts solving relatively easy tasks (with larger $\xi$ value) and is rewarded even for circuits obtaining energies above chemical precision. The sequence of tasks with increasing difficulty helps the agent gradually build knowledge and ultimately create circuits satisfying chemical precision. However, this approach requires the user to design a task sequence which, without \textit{a priori} knowledge of the relationship between the threshold value and the corresponding task difficulty, may lead to the agent reaching undesired local minima.\footnote{We tested multiple variants of this method and the algorithms did not pass the desired threshold with this approach.}
	
	\begin{figure}
		\centering
		\includegraphics[width=.4\textwidth]{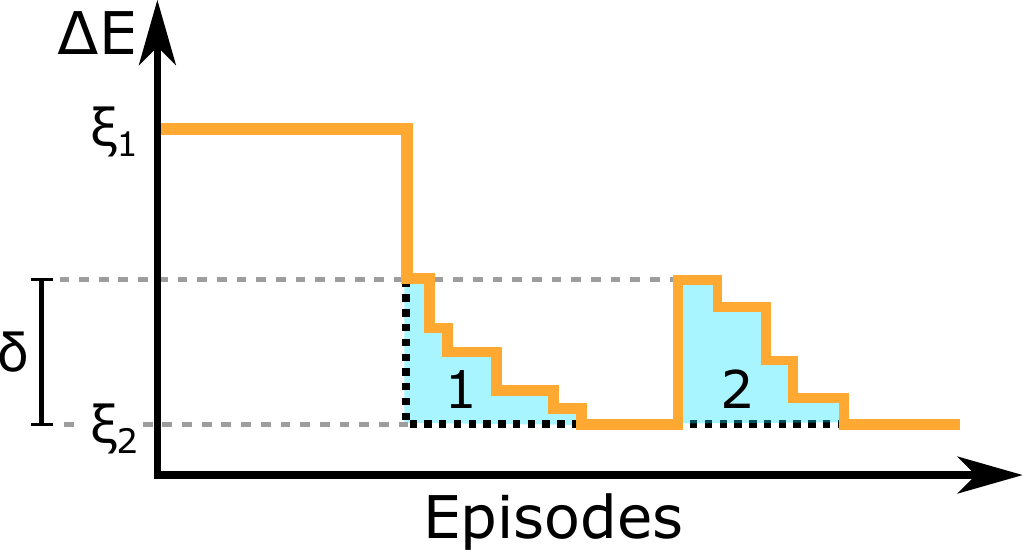}
		\caption{Illustration of the moving threshold (orange) procedure, showing the effects of two amortization events (blue), the value of which is represented by $\delta$. The first event follows a non-zero change in threshold from $\xi_1$ to $\xi_2$, \textit{i.e.} the agent manages to find a better energy estimate during training. The second amortization event illustrates what happens when the agent fails to improve on the current threshold $\xi_2$ (or the improvement is smaller than the amortization value): the threshold increases suddenly due to the resetting of the amortization value. Note that the final threshold after the second amortization drops to zero can also be lower than $\xi_2$.}
		\label{fig:amortization}
	\end{figure}
	
	To solve this issue, inspired by the body of works exploring intrinsic motivation in RL~\cite{aubret2019intrinsic}, we propose the intrinsically motivated curriculum learning in which the difficulty of the current task is dynamically adjusted based on the current performance of the agent. Hence, the agent does not rely on a human-defined task schedule and builds its knowledge gradually at the pace that is reflective of the agent's performance. 
	Namely, the proposed method starts with a threshold level higher than the chemical accuracy. This threshold is then modified by two approaches that work simultaneously.
	In the training process, all energy values obtained by the agent are collected.
	After a fixed number of episodes, the threshold is changed to the level corresponding to the lowest energy value obtained so far, with an additional amortization value. The last factor is to prevent the loss of learning quality due to an overly greedy change of the threshold.
	The second threshold modification approach involves reducing the previously applied depreciation. More specifically, after the first threshold change, a success counter is started which counts the number of episodes that ended with maximal +5 reward. After a fixed number of successes, the amortization value is reduced. We repeat this amortization reduction until the amortization radius is eliminated entirely, \textit{i.e.} the agent finds circuits satisfying the current best threshold. At the same time, the energies obtained in the process are collected and the episodes counted, after which another greedy change of the threshold is performed along with the resetting of the amortization value. This process is illustrated graphically in Fig.~\ref{fig:amortization}.
	As mentioned at the beginning of the section, in the naive approach one can use the exact value of minimum energy $E_{min}$ to compute the rewards. However, due to the relative nature of our reward system we can substitute this value with an arbitrary value $\alpha$ that is lower than minimum energy $E_{min}$ and set a threshold $\xi$ to a value that guarantees that the exact energy falls into the range $ E_{min} \in (\alpha + \xi, \alpha)$. Note that, in principle, the method does not pose any restrictions on the starting threshold and the agent can start learning from any point, so we can set the $\xi$ arbitrarily big to make sure that the $E_{min}$ is in the range. Analogously, the $\alpha$ value can be set, however a simple methods to roughly estimate the lower bound of energy can be utilized to that end. If the agent is able to create circuits with chemical precision relative to any (reasonably) higher energy, the training with every-increasing difficulty will ultimately lead to that point without computing the exact energy.

	\section{Experiments}
	We start this section with description of chemical problems on which we evaluate the proposed approach. Next we will describe experiment setup and present our results. 
	
	\subsection{Experimental setup}
	In our analysis of RL for VQE circuit synthesis we focus on the problem of finding the ground state energy of the lithium hydride (LiH) molecule for various intramolecular distances, as well as Hamiltonians on differing qubit numbers, which stem from different approximations of the true chemical problem.
	All experiments are divided into two parts. In the first part we explain the how the choice of optimizer and number of optimized parameters per step impact the performance of the RL approach. This part is evaluated on less challenging problem instances, i.e. with the smaller quantum system, action space and state space sizes. The second part, on the other hand, will be performed on the same molecule, but simulated on a higher dimensional system. 
	We assume that the number of measurements is sufficiently large that sampling noise is negligible. Whilst this is a significant assumption to make, in this work we are interested in a proof of concept for RL-based circuit synthesis. Thus we do not include the additional overhead incurred by finite measurements, which is left for future work.
	
	\paragraph{Quantum chemistry problems.}
	As mentioned earlier, in this work we focus on finding the ground state energy of the LiH molecule, although our approach design is not specific to this molecule. The Hamiltonians are computed in STO-3G basis. 
	In the first part of our analysis we consider a simpler approximation of LiH molecule which, after taking into account the symmetry of the molecule and removing the orbitals with weak interaction, results in a Hamiltonian defined on 4 qubits. Thus, state space and action space are significantly smaller than the ones for the full LiH Hamiltonian. In this model we utilize the parity mapping to convert molecular Hamiltonian to qubit Hamiltonian \cite{Seeley2012}. We examine proposed method on three values of LiH bond distances, 1.2\r{A}, 2.2\r{A} and 3.4\r{A}.
	In the second experiment we use the larger Hamiltonian which only takes into account the symmetry of the molecule, and is therefore defined on 6 qubits, for which we use the Jordan-Wigner mapping \cite{Jordan1928}. We opted to switch from the parity to the Jordan-Wigner mapping when considering the 6 qubit Hamiltonian in order to compare our results with previous literature which tackled this problem using a different approach \cite{Rattew2019ADN}.
	We only focus one a single geometry for this case, i.e. for bond distance 2.2\r{A}.
	All Hamiltonians were generated using the Qiskit library \cite{Qiskit}.

	\paragraph{Implementation details.}
	In all experiments we utilize $n$-step DDQN algorithm, with the discount factor set to $\gamma =0.88$, and the probability of a random action being selected is set by an $\varepsilon-$greedy policy, with $\varepsilon$ decayed in each step by a factor of $0.99995$ from its initial value 1, down to a minimal value $\varepsilon=0.05$. Memory replay buffer size was set to 20,000.
	The target network in the DDQN training procedure is updated after each 500 actions. After each training episode, we included a testing phase where probability of random actions is set to $\epsilon=0$ and experience replay procedure is turn off, i.e. the experiences obtained during testing phase are not included in the memory replay buffer.

	In our experiments we use differing step sizes in the $n$-step trajectory rollout updates~\cite{bookSuttonBarto2nd}, which specify how exactly the Q-function approximations are updated. This hyperparameter of the model was set to $n=1$ for first part of experiments.
	The reported results correspond to the experiment performed with the value $n=6$. In the moving threshold approach, the threshold is changed greedily after 2000 episodes with amortization radius 0.0001. After 50 successfully solved episodes, amortization is decreased by 0.00001. The initial threshold value is set to $\xi=0.005$. Simulations of quantum circuits were performed using Qulacs library~\cite{suzuki2020qulacs}. The hyperparameters were selected through coarse grain search. 
	
	\paragraph{Evaluation.}
	
	To validate capabilities of reinforcement learning approach we compare it with well established architectures, namely the Hardware Efficient (HE)~\cite{kandala2017he} ansatz and UCCSD~\cite{hoffman1988ucc, bartlett1989ucc} ansatz. The number of layers of the HE ansatz is tailored to each Hamiltonian considered. We report the smallest number of layers for which we have achieved chemical accuracy using our chosen optimizers. Moreover, the naive approach of UCCSD is used, and it is likely possible to find a shallower implementation of this ansatz, but for the purposes of this manuscript, it is an acceptable benchmark. 
	
	In all experiments we compare the minimal depth
	and number of gates of the obtained circuits. By depth we mean the length of the longest path between input and the output along qubit wires, without taking into account quantum gates commutation. Each RL experiment runs on 10 trials, i.e. experiments run on 10 different random seeds. 
	For the first part of experiments we consider two optimization methods: Rotosolve and COBYLA. In turn, each optimization method is applied in with two strategies: \emph{local strategy}, where the optimizer updates the last five rotation gates after each agent step, and \emph{global strategy}, where after each timestep a full update is performed. For the local optimization strategy, we set the number of available iterations of Rotosolve to 5, whilst for the global strategy, the number of iterations for is chosen to be 25. By iterations (for Rotosolve) we mean one complete cycle updating all the parameters under consideration. COBYLA in both strategies will have 100 iterations. 
	Due to the fact that the number of iterations is fixed, convergence is not always guaranteed. However, in most cases, these values were sufficient and more iterations did not improve the energy estimate.
	
	From these experiments, we were able to establish which strategy and which optimiser to use in second part of experiment. The setup for the second part was chosen to be global optimization in each step using COBYLA method with 200 iterations.
	
	\subsection{Results}
	\label{sec:Results}

	\paragraph{Understanding the problem.}

	\begin{figure}
		\centering
		\subfigure{
			\includegraphics[width=0.7\textwidth]{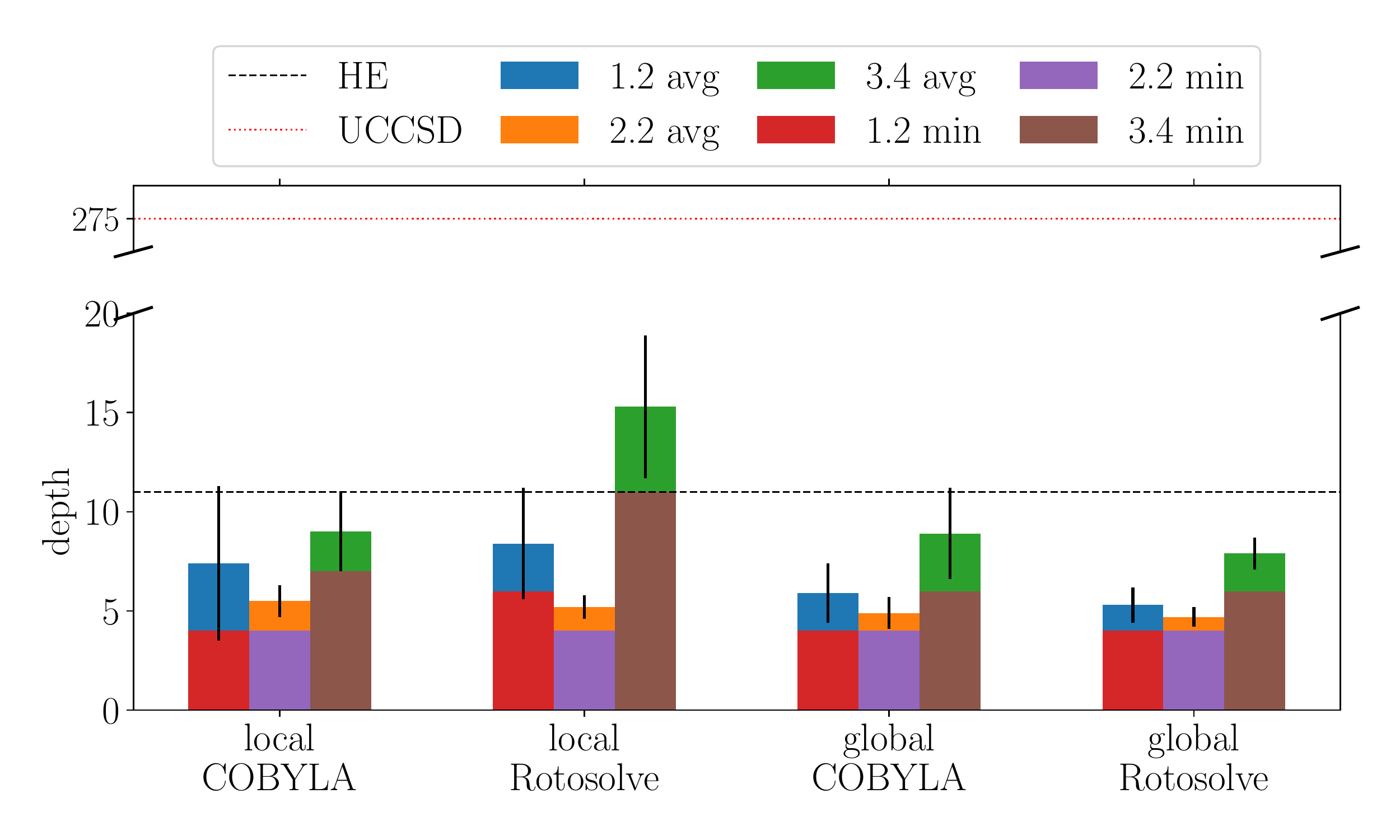}
			\label{fig:histograms_depth}}
		\subfigure{
			\includegraphics[width=0.7\textwidth]{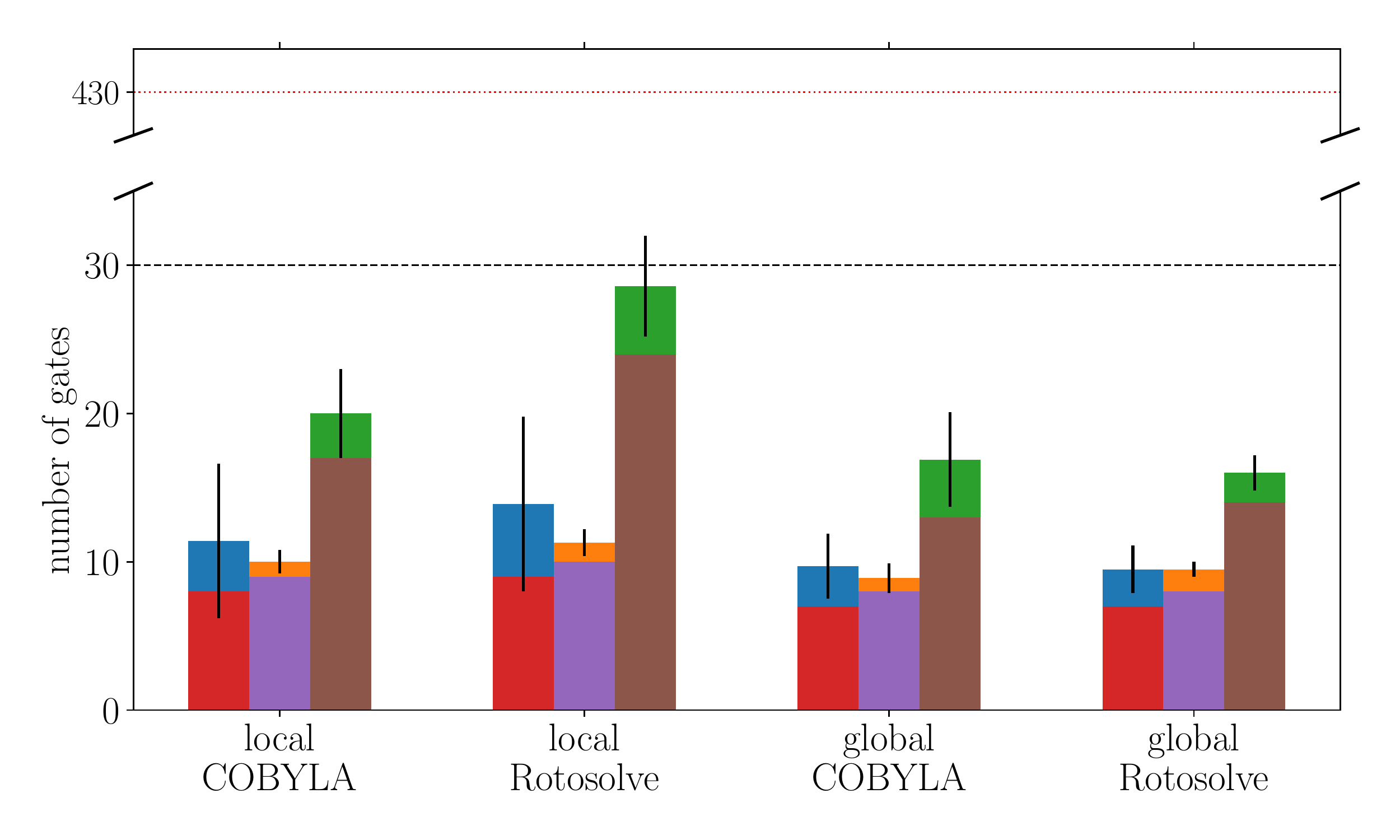}
			\label{fig:histograms_gates}}
		\caption{Comparison of minimum depth (top) and gate count (bottom) of the quantum circuits on which chemical accuracy has been achieved for different strategies. The above results are for the 4 qubit LiH Hamiltonian for several intramolecular separations, shown in the legend.
			The experiments testing the RL approach were run 10 times on different seeds for the networks' parameters and the $\varepsilon$-greedy random actions. Each bar labelled by ``avg'' in the legend represents the average over the results from the different trials, whereas each one labelled by ``min'' is smallest values obtained over the trials.
			For the local optimization strategy with bond distance 3.4\r{A}, the agent found circuits achieving chemical accuracy in two out of ten trials using COBYLA and three out of ten using Rotosolve.}
		\label{fig:hist_4q}
	\end{figure}
	
	In Fig.~\ref{fig:hist_4q}, the depth (top) and the total number of gates (bottom) of the circuits which achieve chemical accuracy are presented, with the average taken over the minimal values from the different trials.
	For bond distances 1.2\r{A} and 2.2\r{A}, the agent proposed quantum circuits which were shallower and contained fewer gates than the standard approaches, in every trial. For bond distance 3.4\r{A}, the agent using local COBYLA optimization found quantum circuits satisfying chemical accuracy in 2 out of 10 trials, whereas the agent using local Rotosolve did so in 3 out of 10 trials. On the other hand, while using global optimization, the agent found circuits satisfying chemical accuracy in every trial, regardless of the optimizer used. 
	In almost all cases, the average minimal depth and gate count are less than those obtained using standard approaches, i.e. HE and UCCSD, the exception being the agent using local Rotosolve, which resulted in circuits deeper than the HE ansatz. 
	Given the above results, COBYLA seems to outperform Rotosolve for this task. Moreover, taking into account number of successful trials, we conclude that our approach gains a lot when all angles are optimized.

	\paragraph{Moving threshold.}

	We evaluated the fixed threshold approach
	on the 6 qubit LiH Hamiltonian case, which, however, did not give any positive results. Agents using global optimization after each step were unable to construct circuits achieving chemical accuracy, which seems to stem from the reward function with chemical accuracy threshold being too sparse. Whilst gradually decreasing the threshold is the obvious next step, we opted for the curriculum learning approach based on intrinsically-driven moving thresholds, in order to decrease the threshold automatically rather than manually.  
	As one can see in Table~\ref{tab:6qubit}, curriculum agents provide better results than standard approaches.
	To the best of our knowledge, the only previous work done tackling this problem for the 6 qubit LiH Hamiltonian is~\cite{Rattew2019ADN}. Whilst we cannot quantitatively compare our results with their work due to differing gate compilations, we note out that our approach seems to generate circuits roughly 5 times shallower than the ones obtained in~\cite{Rattew2019ADN}. 

	\begin{table}[ht!]
		\caption{Comparing the minimum depth and the number of gates obtained using the threshold RL approach with those obtained using the standard HE and UCCSD anst{\"a}ze for circuits which achieve chemical accuracy. The experiments were performed on the 6 qubit LiH molecule with bond distance 2.2\r{A}. For the RL data, ``avg'' denotes the average over minimum depths over different trials, whereas ``min'' denotes the minimum value achieved over all trials. For the standard anst{\"a}ze, the minimum depth and number of gates are obviously fixed by the architectures themselves. The RL approach successfully generated circuits achieving chemical accuracy in 2 out of 10 trials.}
		\label{tab:6qubit}
		\centering
		\begin{tabular}{|l|c|c|c|c|}
			\hline
			& avg depth & min depth & avg \# gates & min \# gates \\ \hline
			RL global COBYLA &  \textbf{14} & \textbf{12} & \textbf{36} & \textbf{29}    \\ \hline
			HE               &  17 & 17 & 63 & 63   \\ \hline
			UCCSD            &   377 & 377 &  610 & 610  \\ \hline
		\end{tabular}
		
	\end{table}

\begin{figure}[ht!]
		\centering
		\includegraphics[width=\textwidth]{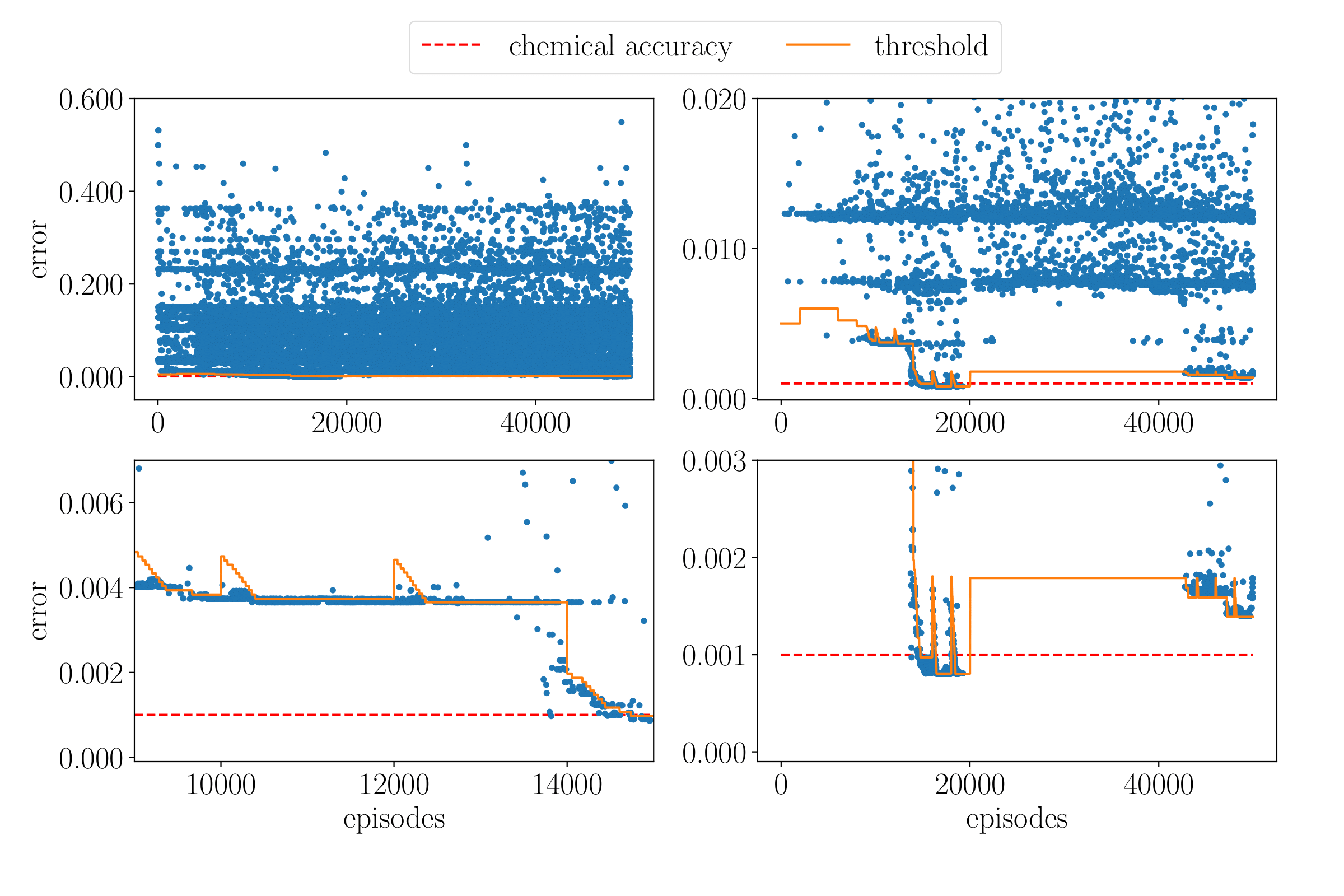}
		\caption{An example of final errors during training, i.e. the difference between minimal energy and energy obtained from the circuits at the end of each episode, with the blue dots representing the error after every episode and the energy threshold, the orange curve showing the energy threshold used to guide the agent, and the red line marking the chemical accuracy. The different plots focus on different episode and error ranges to illustrate the effects of decreasing the threshold, as well as the sudden increases in the latter due to the introduction of amortization values.}
		\label{fig:moving_threshold}
	\end{figure}

	By looking closely at how the moving threshold guides the agent in the direction of chemical accuracy, we can analyze the mean difference between the exact energy and the energy estimate using the agent's circuit from the end of each episode. In Fig.~\ref{fig:moving_threshold} we plot this error for a particular case at different scales to understand better what happens at different training stages. The blue dots representing the error in the agent energy estimates, whilst the orange curve represents the moving threshold. 
	The top left of Fig.~\ref{fig:moving_threshold} shows this error for the entire training period, but does not provide much information due to the large deviation in the error. On the other hand, in the top right plot of Fig.~\ref{fig:moving_threshold}, we focus more on episodes which resulted in errors closer to chemical accuracy. Finally, in the bottom plots in Fig.~\ref{fig:moving_threshold}, we show the behaviour of the agent when guided by a decreasing moving threshold, with the threshold occasionally increasing abruptly due to amortization, which helps the agent adapt to a new lower threshold. 

	\section{Learning procedure with lower-bound approximation to the ground-state
		energy}
	
	So far we have relied on the assumption that the agent uses the energy error with respect to exact solution as a learning metric.
	This assumption in practice is of course not reasonable, as estimating the energy to this precision is a main objective of VQE algorithms to start with.
	One way to relax the assumption of knowing the exact energy to just reasonable information about the lower and upper bounds, and classical and cheaper methods, respectively, and which incurs only a logarithmic overhead in the precision.

	The moving threshold approach discussed in the previous sections not only allows us to solve more demanding problems, like the 6 qubit Hamiltonian, but also removes the requirement of knowing the exact energy. 
	In order to demonstrate this method, we performed an additional experiment on the 4 qubit LiH Hamiltonian at bond distance 2.2\r{A}. Instead of using the exact energy, i.e. $\approx -7.8448$Ha, we used the negative sum of the absolute values of the Hamiltonian's Pauli coefficients, i.e. for a Hamiltonian $H = \sum^M_{j=1} c_j P_j$, we take $-\sum^M_{j=1} |c_j|$ to guide the agent. This is an example of a very rough easy-to-compute lower bound on the energy, but requires the Hamiltonian to be local. For this LiH geometry, this weight sum evaluated to $-10.0604$Ha.
	We set the amortization radius to $0.005$, the number of episodes after greedily shifting the threshold to 500 and the amortization radius is decreased after every 25 successfully solved episodes. The initial threshold value we set to $\xi=4$.  We used global COBYLA with 100 iterations to optimize the angles of the quantum circuit. 
	
	\begin{figure}
		\centering
		\includegraphics[width=\textwidth]{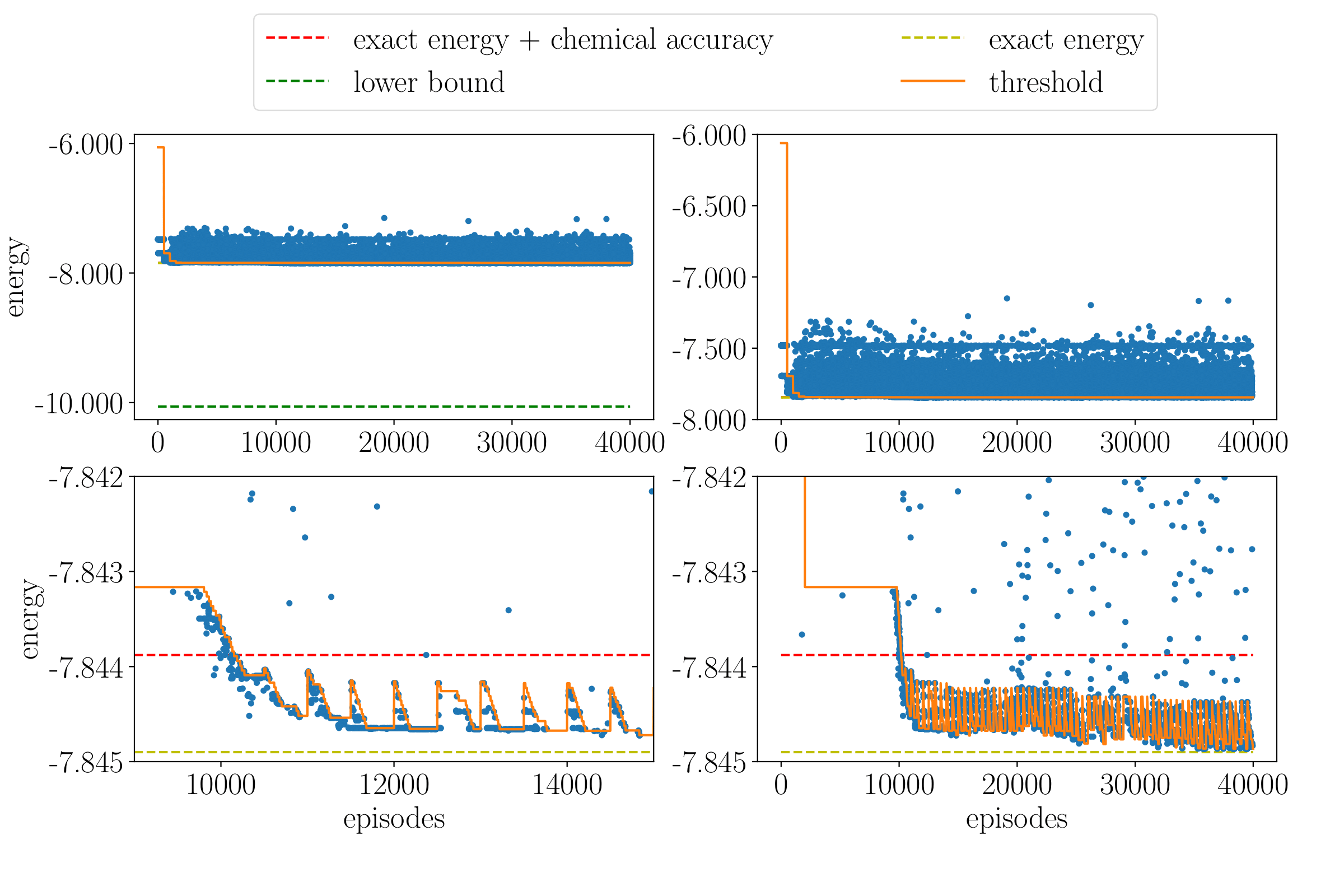}
		\caption{An example of final energies applied to the moving threshold approach without exact energy knowledge. The blue dots represent the energy estimate from the agent's final circuit at the end of each episode, and the orange curve shows the energy threshold used to guide the agent.  The agent manages to reach the exact energy (yellow line) within chemical accuracy (red line) despite the provided estimate (green line) being well below the exact value.}
		\label{fig:fake_energy}
	\end{figure}
	
	As one can see in Fig.~\ref{fig:fake_energy} , the agent successfully manages to reach the exact energy within chemical accuracy, despite having a lower bound below the exact value.
	Moreover, it can be seen that the initial threshold is well above the error that the agent reaches, which also removes the requirement of human input to set the threshold schedule. Similar experiments were performed on bond distances 1.2\r{A} and 3.4\r{A} with the same results.

	\section{Discussion}

	The molecular electronic structure  problem is a promising near-term application for quantum computers. However, the performance of the VQE depends critically on the structure of the chosen ansatz. 
	This ansatz optimization is a combinatorial problem that requires balancing the depth and expressivity of the architecture. Reinforcement learning by design aims at obtaining the shortest optimal solution for a given problem and thus offers a viable approach for structure learning. 
	
	In this work, we present a deep reinforcement learning architecture with intrinsically-motivated curriculum learning to optimize the structure of VQE circuits. The goal of this approach is to design circuits that estimate the ground state energy of molecules within chemical accuracy while keeping the circuit depth as low as possible. 
	In our approach, instead of confronting the RL agent with the full-scale problem from the beginning, the agent first starts training on a simpler instance and autonomously adjusts the task complexity until chemical accuracy is achieved. In particular, we demonstrate that our approach yields the ground state energy of LiH within chemical accuracy for several bond distances. While other VQE approaches also reach chemical accuracy for this benchmark, our approach consistently outperforms the others in terms of circuit depth. Hence, the unique combination of deep RL and intrinsic motivation for structure learning yields an interesting new approach for solving VQE problems. Moreover, we have shown that our approach works using lower-bound approximation to the ground-state
	energy. 
	
	We emphasize that whilst we focus specifically on the molecular electronic structure problem for LiH, the reinforcement learning approach presented in this paper can be adapted for optimizing other VQE architectures. 
	The method is also directly compatible with other reinforcement learning algorithms, and the angle optimization algorithms are not restricted to the methods investigated in this paper. 
	
	The success of curriculum learning  in this paper calls for further exploration and development of such methods.
	Future work might explore the use of similar techniques to transfer knowledge obtained from training on a specific molecular configuration to efficiently explore the space of ans{\"a}tze for different geometries of a given molecule. Investigating the effects of having a finite number measurements on similar RL-based methods would also be an interesting research avenue to pursue in the near future, as NISQ devices are bound by finite measurement numbers. 
	
	Regarding future analysis, it may also be worthwhile to investigate the structures of the circuits returned by a trained agent in more depth. Such an analysis could lead to new insights about the entanglement and other structural properties of the low-energy regions of the underlying systems, and thus shed new light on the the actual physics underneath. Before doing so, pruning the circuits of any method-specific effects may be required first, which may have nothing to do with the physics of the system itself.
	Our preliminary analysis shows that the agent-constructed circuits, although very NISQ-frindly (in the sense of minimized gate numbers and quantum circuit depth), still carry redundancies, e.g., the same gates are sometimes repeated one after another, which are a result of the method, and which could be eliminated.
	This could be achieved, e.g., by using automated postprocessing methods to optimize the circuits (e.g. a Qiskit Terra transpiler~\cites{Qiskit}). The result would be a collection of distinct optimized circuits providing a characterization of the low-energy manifolds.
	
	However, without such circuit optimization/clean-up, we can already identify certain persistent features. For instance, the vast majority of rotations gates used by the agent are $ R_Y $ gates, in all cases we analyzed. At present, it is not obvious whether this is a feature of the method or the systems under study. 
	However, it is difficult to find any additional regularity in the structure of our circuits, beyond such simple observations. In particular, the ansatze we obtain seem very dissimilar to standard architectures. For instance they do not resemble the HE architectures, and they are much shallower than the UCCSD circuits, which makes comparison difficult. A more detailed analysis of the circuits obtained is planned for follow-up work. 
	
	In summary, the methods proposed in this work provide promising initial results, and lay the foundations for the application of deep reinforcement learning methods to VQE optimization problems. And, as is often the case when new methods are employed, we envision numerous follow-up avenues which will reveal the true capacities of such automated ML methods for VQE-type problems.

	\section*{Acknowledgements}
	This manuscript builds on the results presented in the NeurIPS workshop paper~\cites{Ost2019nips}.
	MO acknowledge the support of the Foundation for Polish Science (FNP) under grant number POIR.04.04.00-00-17C1/18-00.
	LMT acknowledges the support
	from the Austrian Science Fund (FWF) through
	the projects DK-ALM:W1259-N27 and SFB BeyondC
	F7102. 
	This work was also supported by the Dutch Research Council(NWO/OCW), as part of the Quantum Software Consortium programme (project number 024.003.037).
	VD and ES acknowledges the support of SURFsara through the QC4QC project.
	This research was partially funded by the Grant of Priority Research Domain at Warsaw University of Technology - Artificial Intelligence and Robotics. The authors also wanted to thank Arthur G. Rattew for helpful discussions. 
	\printbibliography
\end{document}